\RequirePackage{fix-cm}
\documentclass[smallextended]{svjour3}       
\smartqed  % flush right qed marks, e.g. at end of proof

\usepackage{graphicx}
\usepackage{graphicx}
\usepackage{color}
\usepackage{xspace}

\usepackage{graphicx,float}
\usepackage{epsf}
\usepackage{bm}
\usepackage{amsmath,hyperref}
\usepackage{amsfonts}
\usepackage{amssymb}
\usepackage{epstopdf}
\usepackage{rotating}
\usepackage{pdflscape}
\usepackage{adjustbox}

\def\laq{~\raise 0.4ex\hbox{$<$}\kern -0.8em\lower 0.62ex\hbox{$\sim$}~}
\def\gaq{~\raise 0.4ex\hbox{$>$}\kern -0.7em\lower 0.62ex\hbox{$\sim$}~}

\def\beq{\begin{equation}}
\def\eeq{\end{equation}}
\def\bea{\begin{eqnarray}}
\def\eea{\end{eqnarray}}

\def \pa {\partial}
\def \ra {\rightarrow}
\def \ti {\widetilde}

\def \Mp {M_{\rm P}}
\def \La {\Lambda}
\def \Da {\Delta}
\def \da {\delta}
\def \b {\beta}
\def \a {\alpha}

\def \Ga {\Gamma}
\def \ga {\gamma}

\def \da {\delta}

\def \Om {\Omega}

\begin{document}

\title{Comparing the luminosity distance for gravitational waves 
and electromagnetic signals in a simple model of quadratic gravity}


\titlerunning{Luminosity distance in a model of quadratic gravity}        
% if too long for running head

\author{G. Fanizza \and G. Franchini \and \\M. Gasperini \and L. Tedesco}

\authorrunning{Fanizza, Franchini, Gasperini and Tedesco} 
% if too long for running head

\institute{
G. Fanizza \at
Instituto de Astrofis\'ica e Ci\^encias do Espa\c{c}o,
Faculdade de Ci\^encias da Universidade de Lisboa,
Edificio C8, Campo Grande, P-1740-016, Lisbon, Portugal \\
\email{gfanizza@fc.ul.pt} 
\and
G. Franchini \at
Dipartimento di Fisica, Universit\`a di Pisa, Largo B. Pontecorvo 3, 56127 Pisa, Italy,\\
and Istituto Nazionale di Fisica Nucleare, Sezione di Pisa, Italy\\
\email{giovanni.franchini@phd.unipi.it}       
\and
M. Gasperini \at
Dipartimento di Fisica, Universit\`a di Bari, 
Via G. Amendola 173, 70126 Bari, Italy,\\
and Istituto Nazionale di Fisica Nucleare, Sezione di Bari, Italy\\
\email{gasperini@ba.infn.it}
 \and
L. Tedesco \at
Dipartimento di Fisica, Universit\`a di Bari, 
Via G. Amendola 173, 70126 Bari, Italy,\\
and Istituto Nazionale di Fisica Nucleare, Sezione di Bari, Italy\\
\email{luigi.tedesco@ba.infn.it}
}

\date{Received: date / Accepted: date}

\maketitle

\begin{abstract}
We compute the modified friction coefficient controlling the propagation of tensor metric perturbations in the context of a generalized  
cosmological scenario based on a theory of gravity with quadratic curvature corrections. In such a context we discuss the differences between gravitational and electromagnetic luminosity distance, as well as the differences with the standard results based on the Einstein equations. We present numerical estimates of the modified luminosity distance on the cosmic redshift scale typical of Supernovae and standard sirens.

 %\subclass{MSC code1 \and MSC code2 \and more}
\end{abstract}

\vspace{1cm}

%% For one-column wide figures use
%\begin{figure}
%% Use the relevant command to insert your figure file.
%% For example, with the graphicx package use
%  \includegraphics{example.eps}
%\end{figure}
%%

%% For two-column wide figures use
%\begin{figure*}
%% Use the relevant command to insert your figure file.
%% For example, with the graphicx package use
%  \includegraphics[width=0.75\textwidth]{example.eps}
%\end{figure*}
%%

%\begin{table}

%\begin{tabular}{lll}
%\hline\noalign{\smallskip}
%\end{tabular}
%\end{table}

%%%%%%%%%%%%%%%%%%%%%%%%%
% INIZIO TESTO LAVORO
%%%%%%%%%%%%%%%%%%%%%%%%%

\section {Introduction}
\label{sec1}

It is well known that measuring the relative luminosity distance of different astrophysical sources, or the relative time-of-flight of different ultra-relativistic particles emitted by the same source (see e.g. \cite{1}), we can obtain important information and constraints on the fundamental cosmological parameters. Recently, the simultaneous detection of gravitational wave (GW) and electromagnetic (e.m.) signals with the same origin -- namely, the binary neutron star merger GW170817 \cite{abbott6} and the associate $\ga$-ray burst GRB170817A \cite{Goldstein:2017mmi}-\cite{GBM:2017lvd} -- has opened the possibility of including also GW observations among the useful tools for testing the standard $\La$CDM model and its possible modifications. Particularly useful (as noted long ago \cite{Schutz:1986gp}) to that purpose are  GW sources like coalescing compact binaries located at different cosmic distances, which can play the role of ``standard sirens" analogous to the role of the ``standard candles" typical of e.m. signals.

In such a context it is important to note that, by comparing the propagation of GW signals received from a given source and the propagation of other ultra-relativistic signals emitted by the same source, we have also the possibility of testing alternative theories of gravity on a cosmic scale of distance, as recently stressed and discussed in 
\cite{Belgacem:2017ihm,Belgacem:2018lbp,B1,11}
(see also \cite{12,13,14,15,15a}).

We should recall, in this respect, that in the standard Einstein's theory of gravity both GW and e.m. signals propagate with the same speed along the light cone of the given background geometry, thus defining exactly the same luminosity distance $d_L$ for the emitting source. In models of gravity different from General Relativity, on the contrary, the propagation of GW and e.m. signals may be different, and not only because the metric satisfies modified Einstein equations, whose solution may define a background geometry different from the standard one. In addition, in fact, the propagation of massless signals may deviate from the usual light-cone propagation, and the deviation of GW signals may be different from the deviation of e.m. signals.

Such a modified propagation may be associated, in particular,  to  a different speed and/or to a different variation of the amplitude of the signal with the distance from the source (for what concerns a possible modified speed of GW signals, however, there are strong constraints imposed by present observations \cite{Monitor:2017mdv}, see also \cite{Creminelli:2017sry,Sakstein:2017xjx,Ezquiaga:2017ekz,Baker:2017hug}). The modified variation of the amplitude, in its turn, produces an effective variation of the associated luminosity distance. Hence,  when computing the relevant luminosity distance in the context of a generalized theory of gravity, we may have to face the (in principle interesting) situation in which $d_L^{GW} \not= d_L^{em} \not=d_L^{GR}$, where the last value of $d_L$ refers to the one computed for the  cosmological scenario based on the General Relativity equations. Comparing the above  (possibly different) three expressions of $d_L$ may thus represent an efficient tool for testing alternative theories of gravity on cosmological scales of distance \cite{Belgacem:2017ihm,Belgacem:2018lbp,B1,11}.

In this paper we will concentrate on a simple generalization of the Einstein theory which contains only one additional (constant) parameter, controlling a quadratic curvature correction to the Einstein action. The model is also known as ``Starobinsky gravity" \cite{Sta}, and can be regarded as a truncated higher-curvature expansion of the gravitational action.  Also, it corresponds to a simple type of $f(R)$-gravity and can be related to the largest class of Horndeski theories, widely used to test gravity through modified GW propagation (see in particular \cite{21} and \cite{11,13,14}).
 It was shown \cite{cosmai} that such a model can fit the Hubble diagram of Supernovae, consistently with other recent cosmic data, even in the absence of an explicit dark-energy contribution. 
Here we show that such a simple modification of Einstein's gravity is already enough to produce an interesting difference between the propagation of e.m. and GW signals, and to predict possibly testable effects when estimating the luminosity distance of standard sirens. 

The paper is organized as follows. In Sect. \ref{sec2} we recall some general result concerning the propagation of tensor metric perturbations in modified theories of gravity and the corresponding definition of the modified luminosity distance. In Sect. \ref{sec3} we apply such result to a gravitational action with quadratic curvature corrections. 
In Sect. \ref{sec4}, by assuming that the deviations from the Einstein theory are small, we obtain an approximate expression for the GW luminosity distance by expanding to first order the solution of the modified cosmological equations around the standard-model solution. In Sect. \ref{sec5} we numerically evaluate the parameters of such a modified cosmological solution by fitting the  UNION 2 data set \cite{amulla} of Supernovae. 
In Sect. \ref{sec5a} we compare our results with the constraints imposed by the standard siren GW170817 and its e.m. partner GRB170817A. 
In Sect. \ref{sec5b} we briefly discuss the differences between the GW and e.m. luminosity distance and the standard luminosity distance $d_L^{GR}$, based on the Einstein equations and on exact light-cone propagation. Finally, Sect. \ref{sec6} is devoted to our conclusive remarks. In the Appendix we present an explicit computation of the propagation equation for tensor metric perturbations for  our model with higher-curvature corrections. 

%%%%%%%%%%%%%%%%%%%%%%%%%%%%%%%%%%%%%%
%%%%%%%%%%%%%%%%%%%%%%%%%%%%%%%%%%%%%%

%%%%%%%%%%%%%%%%%%%%%%%%%%%%%%%%%%%%%%
%%%%%%%%%%%%%%%%%%%%%%%%%%%%%%%%%%%%%%

\section {GW luminosity distance in generalized models of gravity}
\label{sec2}

Let us start by recalling, for the reader's convenience, a few well know results concerning the propagation of tensor metric perturbations in a spatially flat, four-dimensional geometry of the Friedmann-Lemaitre-Robertson-Walker (FLRW) type, described in conformal time $\eta$ by the background metric $g_{00}= a^2(\eta)$, $g_{ij}= -a^2(\eta) \da_{ij}$. In the so-called transverse-traceless gauge, each polarization component $h$ of the first order metric perturbation $h_{\mu\nu}$ satisfies the linearized propagation equation
\beq
{d^2h\over d\eta^2}+ 2 {\cal H}{d h\over d\eta}-\nabla^{2}h=0,
\label{21}
\eeq
where we have defined ${\cal H}=a^{-1}da/d\eta$ (we have assumed no anisotropic stress in the matter sources). The corresponding canonical variable \cite{Muk,25a} $v=ah$ satisfies
\beq
{d^2v\over d\eta^2}-\left(\nabla^{2}+{1\over a}{d^2 a\over d\eta^2}\right)v=0.
\label{22}
\eeq

By expanding the solution in Fourier modes, let us consider those modes well ``inside" our present horizon, satisfying the condition $\left|a^{-1} d^2a /d\eta^2 v \right| \ll \left|\nabla^2 v\right|$, like (in particular) the modes relevant to the sensitivity frequency band of present interferometric detectors. For such modes the equation for $v$ is unaffected by the background geometry, so that the amplitude of a gravitational signal $h=v/a$ emitted at the time $\eta$ and received at the time $\eta_0>\eta$ decreases, during the propagation from the source to the observer, by an amount determined by the standard redshift factor, i.e. by the factor $a(\eta)/a_0 \equiv (1+z)^{-1}$, where $a_0\equiv a(\eta_0)$, and  $z$ is the redshift of the emitting source.

On the other hand, if we are interested in GW signal emitted by coalescing binaries (like the so-called standard sirens), it is well known \cite{maggiorelibro1,maggiore2}   that such a ``weakening" of the amplitude (due to the propagation across the expanding cosmic geometry) leads to introduce the so-called luminosity distance of the emitting source defined in terms of the received flux of radiation, and thus inversely proportional to the amplitude $h$ of the received signal, namely 
$h \sim d_L^{-1}(z)$. Here $d_L$ is the standard luminosity distance referred to signals propagating on the light-cone of our FLRW geometry, and given, for a metric with scale factor $a$ and Hubble factor $H= a^{-2} da/d\eta$, by
\beq
d_{L}(z) = (1+z)\int^z_0 \frac {dz'} {H(z')}
\label{23}
\eeq
(see also \cite{Belgacem:2017ihm,Belgacem:2018lbp,B1,11} and \cite{13,14} for a detailed discussion of how the GW amplitude $h$ is related to the GW luminosity distance).

Let us now consider, following \cite{Belgacem:2017ihm,Belgacem:2018lbp,B1,11}, a generalized model of gravity different from Einstein's theory and leading, in particular, to a different set of cosmological equations. Their perturbation gives us a modified equation for the tensor modes $h$ which, even in the same FLRW metric as before, takes a form different from Eq. (\ref{21}):
\beq
{d^2h\over d\eta^2}+ 2 {\cal H}{d h\over d\eta} \left[1- \delta(\eta)\right] -c_s^2(\eta) \nabla^{2}h=0.
\label{24}
\eeq
We may have, in general, a modified speed parameter, $c_s\not=1$, and a modified geometric friction term with $\da \not=0$. Both parameters $c_s$ and $\da$ may be in principle time-dependent. Even in that case, however, we can still introduce the analogous of a canonical variable $v$ defined by $v\equiv \ti a h$, where $\ti a$ satisfies
\beq
{1\over\ti{a}}
\frac{d\,\ti{a}}{d\eta}={\cal H}\left[1-\delta(\eta)\right],
\label{25}
\eeq
and obtain the corresponding propagation equation in canonical form (but with a modified ``pump field", $a \ra \ti a$) as follows:
\beq
{d^2v\over d\eta^2}-\left(c_s^2\,\nabla^{2}+{1\over \ti a}{d^2 \ti a\over d\eta^2}\right)v=0.
\label{26}
\eeq

Let us concentrate, again, on modes well inside the horizon, and on a theory of gravity which implies $c_s=1$. In that case, the only difference from the previous case described by the perturbation of the Einstein equations is that the amplitude $h= v/\ti a$ of the GW signal, while traveling from the source to the observer, is now decreased (or increased) by the modified factor $\ti a(\eta)/\ti a_0 \equiv (1+\ti z)^{-1}$ (instead of the standard redshift factor $a(\eta)/a_0$). Hence, the received amplitude turns out to be inversely proportional to a modified luminosity distance, which we may call $d_L^{GW}$, and which we may relate to the standard luminosity distance $d_L$ as follows \cite{Belgacem:2017ihm,Belgacem:2018lbp,B1,11}: 
\bea
{1\over d_L^{GW}(z)}&=& \left(1+z\over 1+ \ti z\right) {1\over d_L(z)}\equiv
{\ti a (z)\over \ti a_0} {a_0\over a(z)}{1\over d_L(z)}=
{\ti a(z)\over a(z)} {1\over d_L(z)}
\label{27}
\eea
Here $d_L$ is the distance given by Eq. (\ref{23}) and associated with the standard light-cone propagation in the geometry with scale factor $a$. Also, we have normalized the solution of Eq. (\ref{25}) for $\ti a$ in such a way that $\ti a(\eta_0) = a(\eta_0)$, so as to recover the obvious result $d_L^{GW}(0)=d_L(0)$ for a source at zero cosmic distance. With such a normalization, a simple integration of Eq. (\ref{25}), rewritten as $d(\log a/\ti a)/d\eta=  a^{-1}(da/d\eta) \da$, leads to express $\ti a$ in terms of $a$ as follows \cite{Belgacem:2017ihm,Belgacem:2018lbp,B1}:
\beq
{a\over \ti a}(z) = {d_L^{GW}\over d_L}(z)= \exp\left[ -\int_0^z {dz'\over 1+z'}\da(z')\right].
\label{28}
\eeq

In the following sections we will apply the above results to a generalized model of gravity with quadratic curvature corrections.

%%%%%%%%%%%%%%%%%%%%%%%%%%%%%%%%%%%%%%
%%%%%%%%%%%%%%%%%%%%%%%%%%%%%%%%%%%%%%

%%%%%%%%%%%%%%%%%%%%%%%%%%%%%%%%%%%%%%
%%%%%%%%%%%%%%%%%%%%%%%%%%%%%%%%%%%%%%

\section {A simple model of quadratic gravity}
\label{sec3}

The generalized model of gravity we shall consider here is a particularly simple case of the so-called $f(R)$-theories (see e.g. \cite{FR6,FR8,Clifton:2011jh} for comprehensive reviews), with a function $f(R)=R+c_2R^2$, which we may also regard as an expansion in power series of the scalar curvature $R$, truncated to second order. Our generalized action is then
\beq
S= -{\Mp^2\over 2 } \int d^4 x \sqrt{-g} \left( R+ c_2 R^2 \right) + S_m,
\label{31}
\eeq
where $S_m$ is the matter action (assumed to describe perfect fluid sources), $\Mp=1/\sqrt{8\pi G}$ is the Planck mass, and $c_2$ an unknown constant parameter with dimensions of length squared (we are using units $\hbar=c=1$). 

By varying with respect to the metric $g_{\mu\nu}$, and by adding to the action (\ref{31}) an appropriate generalized version of the York-Gibbons-Hawking action (see e.g. \cite{cosmai}), so as to get rid of the unwanted boundary contributions (as in General Relativity), we obtain the following generalized Einstein equations
\bea
\!\!\!\!\!\!\!\!\!\!
&&
8 \pi G \,T_\mu^\nu=R_\mu^\nu -{1\over 2} \da_\mu^\nu R 
+c_2 \left[2RR_\mu^\nu+  \da_\mu^\nu\left(2 \Box R-{1\over 2} R^2 \right) -2 \nabla_\mu \nabla^\nu R\right],
\label{32}
\eea
which are fourth-order differential equations for the components of the metric tensor. We have defined $\Box \equiv \nabla_\mu\nabla^\mu$, and $T_{\mu\nu}$ is the source stress tensor, given by the variation with respect to $g_{\mu\nu}$ of the matter part of the action. It can be easily checked that by taking the covariant divergence of both sides of the above equation, and using the contracted Bianchi identity, one consistently recovers the standard result
\beq
\nabla_\nu T_\mu\,^\nu=0,
\label{32a}
\eeq
which guarantees the covariant conservation of the stress tensor of the gravitational sources.

Let us specify now the above equations to the case of a spatially flat FLRW metric with scale factor $a$, denoting with a dot the derivative with respect to the  cosmic time coordinate $t$, related to conformal time by $dt=ad\eta$. We shall adopt the following conventions: $g_{\mu\nu}= {\rm diag} (+---)$, $R_{\nu\a}= R_{\mu\nu\a}\,^\mu$, and $R_{\mu\nu\a}\,^\b= \pa_\mu \Ga_{\nu\a}\,^\b- \cdots$. We then find, in particular, 
\beq
R_0^0=-3\left(\dot H +H^2\right), ~~~ R_i^j= -\left(\dot H +3H^2\right)\da_i^j,
\label{33}
\eeq
and
\beq
R=-6\left(\dot H +2H^2\right),
\label{34}
\eeq
where $H=\dot a/a$. For a perfect fluid source, with $T_\mu^\nu= {\rm diag}(\rho, -p\da_i^j)$, the time component of Eq. (\ref{32}) thus gives
\beq
3H^{2} +18 c_{2}\left[\dot{H}^{2}-6H^{2}\dot{H}-2H\ddot{H}\right]= 8 \pi G \rho.
\label{35}
\eeq
The space-like (diagonal) components of Eq. (\ref{32}) lead to the condition
\beq
2\dot{H}+3H^{2} \!-6c_{2}\!\left[9\dot{H}^2\!+\!18\dot{H}H^{2}+12H\ddot{H}\!+\!
2\dddot{H}\right]\!=\!- 8 \pi Gp.
\label{36}
\eeq
The mixed components of Eq. (\ref{32}) are identically satisfied. Finally, the conservation equation (\ref{32a}) takes the usual form
\beq
\dot \rho + 3H(\rho+p)=0.
\label{36a}
\eeq

To obtain the evolution equation for the first-order tensor perturbations of the metric, defined by $g_{\mu\nu} \ra g_{\mu\nu}+\da g_{\mu\nu}$, $\da g_{\mu\nu}=h_{\mu\nu}$, we shall work in the transverse-traceless (TT) gauge $\nabla_\nu h_\mu^\nu=0=h_\mu\,^\mu$, which in our context takes the simple form $h_{\mu0}=0$, $\nabla_jh_i\,^j\equiv \pa_j h_i\,^j=0=g^{ij}h_{ij}$. By directly perturbing the background equations  (\ref{32}) (see Appendix A for an explicit computation) we then obtain for each polarization mode of $h_i\,^j$ a linearized equation which, expressed in conformal time, can be written as
\beq
{d^2 h\over d\eta^2}+2{\cal H}\left[1+ {c_2 (dR/d\eta)\over {\cal H}(1+2c_2R)}\right]{dh\over d\eta}-\nabla^2 h=0,
\label{37}
\eeq
where $R$ is the scalar curvature (\ref{34}). Comparing with the modified equation (\ref{24}) we then find, for our model, a standard propagation velocity $c_s=1$, but a non-standard friction term which can be written in the form (\ref{24}) with the time-dependent parameter
\beq
\da(\eta)=-  {c_2 (dR/d\eta)\over {\cal H}(1+2c_2R)}.
\label{38}
\eeq

It is important to recall, at this point, that the absence of modifications of the standard speed parameter is a peculiar property of the model of gravity we are considering. More general quadratic-curvature corrections to the Einstein-Hilbert action can lead indeed to a propagation equation (\ref{24}) with an effective speed parameter $c_s \not=1$. An important example of this type is provided by the Gauss-Bonnet two-form typically  predicted in a string-theory context \cite{32}, and  generated by the high-curvature $\a'$ corrections to the String-frame gravi-dilaton action (see \cite{gasperini2} for an explicit computation of the modified propagation of tensor perturbations in the presence of the quadratic Gauss-Bonnet term in the gravi-dilaton action). 

It may be useful to check, also, that the previous results (\ref{37}), (\ref{38}) can be exactly reproduced starting with the modified action (\ref{31}), and expanding the action up to terms quadratic in the metric fluctuations $h_i\,^j$ (so as to obtain an effective Lagrangian for the first-order perturbations, from which to deduce their equations of motion through the standard variational formalism). 

In fact, working again in the TT gauge, we have the following quadratic contributions to the perturbed action (see also \cite{gasperini2,gasp,gasperini}):
\beq
\!\!\!\!\!\!\!\!\!\!\!\!\!\!\!\!\!\!\!\!\!\!\!\!\!\!\!\!
\delta^{(2)}\sqrt{-g}=-{1\over 4} \sqrt{-g}\, {\rm Tr} \,(h^2),
\label{39}
\eeq
\bea
 \delta^{(2)}(\sqrt{-g}R)&=&a^{3}{\rm Tr} \left[\left(\frac{3}{2}\dot{H}+3H^{2}\right)h^{2}+h\ddot{h} \, + \right.
 \nonumber\\ &+& 
 \left.
4H h \dot{h}+\frac{3}{4}\dot{h}^{2}-\frac{1}{4}h\frac{\nabla^{2}}{a^{2}}h\right],
 \label{310}
 \eea
 \bea
 \delta^{(2)}(\sqrt{-g}R^{2})&=&R \, a^{3}{\rm Tr} \left[\left(\frac{3}{2}\dot{H}+3H^{2}\right)h^{2}+2h\ddot{h} \, + \right.
 \nonumber\\ &+& \left.
 8H h \dot{h}+\frac{3}{2}\dot{h}^{2}-\frac{1}{2}h\frac{\nabla^{2}}{a^{2}}h\right],
\label{311}
\eea
where ${\rm Tr}(h^{2})={h_{i}}^{j}{h_{j}}^{i}$, ${\rm Tr}(h\dot{h})={h_{i}}^{j}\dot{ h_{j}}^{i}$, and so on. Summing up all contributions, integrating by parts the terms in 
$h\dot{h}$ and $h\ddot{h}$, and neglecting a total derivative which does not contribute to the equations of motion, it turns out that the coefficient of the $h^2$ term appearing in the quadratic perturbed action is identically vanishing, thanks to the background equation (\ref{36}). By decomposing the fluctuations into the two physical polarization modes $h_+$ and $h_\times$, i.e. ${\rm Tr}(h^{2})={h_{i}}^{j}{h_{j}}^{i}= 2 (h_+^2 + h_\times^2)$, and switching again to the confromal time coordinate where $\dot h= a^{-1}dh/d\eta$, we are eventually led (for each polarization mode) to the following quadratic effective action:
\beq
\delta^{(2)}S=\frac{\Mp^2}{4}\int d^{3}xd\eta \,a^{2}\left(1+2c_{2}R\right)\left[\left(dh\over d\eta\right)^{2}+h\nabla^{2}h\right].
\label{312}
\eeq

We can now easily recover the standard form of the Einstein perturbed action by defining a modified scale factor $\ti a$, related to $a$ (and satisfying the correct normalization $\ti a_0=a_0$, see Sect. \ref{sec2}) in such a way that
\beq
\ti a(\eta)= a(\eta) \left[ 1+2c_2 R(\eta)\over 1+2 c_2 R(0)\right]^{1/2}.
\label{313}
\eeq
By replacing $a$ with $\ti a$ in the action (\ref{312}), and applying to $h$ the standard variational formalism, we are then led to the equation of motion   for tensor perturbations in the form
\beq
{d^2 h\over d\eta^2}+  {2\over \ti a}{d \ti a \over d\eta}{dh\over d\eta}-\nabla^{2}h=0.
\label{314}
\eeq
Finally, by computing $\ti a'/\ti a$ from the definition (\ref{313}), and comparing the result with Eq. (\ref{37}), we can immediately recover the expression of the parameter $\da (\eta)$ anticipated in Eq. (\ref{38}).

To conclude this Section let us stress that by following this second procedure, based on the computation of the quadratic perturbed action, we can directly obtain from Eq. (\ref{313}) the important ratio $d_L/d_L^{GW}= \ti a/a$ (see Eq. (\ref{28})),  with no need of performing the  explicit integration of the parameter $\da(\eta)$ along the travel of the signal from the source to the observer.

%%%%%%%%%%%%%%%%%%%%%%%%%%%%%%%%%%%%%%
%%%%%%%%%%%%%%%%%%%%%%%%%%%%%%%%%%%%%%

%%%%%%%%%%%%%%%%%%%%%%%%%%%%%%%%%%%%%%
%%%%%%%%%%%%%%%%%%%%%%%%%%%%%%%%%%%%%%

\section {First-order corrections to the standard cosmological dynamics}
\label{sec4}

In the previous Section we have introduced a modified theory of gravity where the propagation of GW signals, and the associated luminosity distance $d_L^{GW}$, are different from those predicted by the standard Einstein theory. In the following sections we will discuss the possibility that such a modified GW dynamics may be compatible with (and possibly useful to interpret) present cosmological observations, by concentrating our attention, in particular, on the distance scales typical of the  Supernovae and of the GW sources known as standard sirens.

In order to compare the model predictions with the observational data in that range of distances, what we need, first of all, is to express both the friction correction $\da$ and the ratio $\ti a/a=d_L/d_L^{GW}$ as a function of the redshift parameter $z$.
To this purpose, by recalling the definition $(1+z)=a_0/a$, we first note that $\dot z= - (1+z)H$, so that
\beq
{d R\over d\eta}= a \dot R= a \dot z{dR\over dz} = -(1+z) {\cal H} {dR\over dz}.
\label{40}
\eeq
We shall adopt, from now on, the convenient notation in which a prime denotes differentiation with respect to $z$. We can then rewrite Eqs. (\ref{38}) and (\ref{313}), respectively, as
\bea
&&
\delta(z)=c_{2}\frac{(1+z)R'}{1+2c_{2}R},
\label{41} \\ && \!\!\!\!
\frac{d_{L}^{GW}}{d_{L}}(z)=\left[1+2c_{2}R(0)\over1+2c_{2}R(z)\right]^{1/2},
\label{42}
\eea
where, from Eq. (\ref{34}),
\bea
&&
\!\!\!\!
R(z)=-6H\left[2H-(1+z)H'\right],
\label{43} \\ && \!\!\!\!
R'(z)=-6\left[3HH'-(1+z)\left(H^{\prime 2} + H H''\right)\right].
\label{44}
\eea
It may be noted that, in spite of the fact that the exact definition of the modified luminosity distance (\ref{42}) is formally equivalent to the GW luminosity distance defined in the presence of a time-varying effective Planck mass (considered e.g. in \cite{11}), the particular $z$-dependence predicted by the above equations  is quite different, in principle, from the $z$-dependence of $d_L^{GW}$ assumed in \cite{11} for their phenomenological discussion.

For an explicit evaluation of the modified luminosity distance (\ref{42})  we need now $H(z)$, which has to be determined by solving the modified cosmological equations (\ref{35}), (\ref{36}). In addition, and for a numerical estimate of the differences with the standard results, we should determine a possible allowed range of values for the parameter $c_2$. To this second purpose we will follow the procedure already suggested in a previous paper \cite{cosmai}, namely fitting a set of Supernovae data with an expression of the luminosity distance $d_L$ computed according to the standard definition (\ref{23}), but with an Hubble parameter $H(z)$ which satisfies the modified cosmological equations of this paper. 

It is important to stress, in this respect, that in spite of the modified gravitational action, the matter part of our model (\ref{31}) keeps unchanged. This implies, in particular, that the e.m. field is still minimally coupled to the  background geometry  (like in General Relativity), and that the propagation equation of  e.m. signals keeps the same form as in the standard Einstein theory. Hence, the e.m. luminosity distance, $d_L^{em}$, is still given by Eq. (\ref{23}), with the only difference that $H(z)$ satisfies the generalized cosmological equations (\ref{35}), (\ref{36}) with $c_2 \not=0$. A fit of the Supernovae data with such a generalized form of $d_L(z)$ can thus provide a reliable estimate of the possibly allowed value of $c_2$. 

In order to obtain the required solution for $H(z)$ we shall work under the assumption that the higher-curvature corrections to the standard equations are small, and that the solution of the modified equations can be expanded in power series of $c_2$ around the zeroth-order solution of the standard cosmological equations. 
 
 Let us consider, in particular, Eqs. (\ref{35}), (\ref{36a}) as our independent equations, and use the standard ($\La$CDM)  model of perfect fluid source with two non-interacting components, namely $\rho= \rho_m+\rho_\La$, with $p_m=0$ and $p_\La= - \rho_\La$.  The conservation equation (\ref{36a}), expressed in terms of $z$, thus reduces to
 \beq
\rho_\La'=0~, ~~~~~~~~~~ (1+z) \rho_m' =3 \rho_m ~,
\label{44a}
\eeq
and can be solved exactly to give $\rho_\La= \rho_{\La 0}$ and $\rho_m= \rho_{m 0} (1+z)^3$, where $\rho_{\La 0}$ and $ \rho_{m 0}$ are integration constants.
By inserting such solution into the generalized Friedmann equation (\ref{35}) we then find that the exact equation for $H(z)$ takes the form:
\bea
&&
\!\!\!\!\!
3H^2+ 18 c_2 H^2 \left[ 4(1+z) H H' - (1+z)^2 (H^{\prime 2} +2H H'') \right]=
\nonumber \\ && =
8 \pi G \left[\rho_{m 0} (1+z)^3 + \rho_\La \right].
\label{44b}
\eea

Let us now expand the solution of the above equation, to first order in $c_2$, as follows:
 \beq 
H(z)= H^{(0)}(z)+ c_2 H^{(1)}(z)+ \cdots, 
\label{44c}
\eeq
where $H^{(0)}$ is the solution of the standard  equation with $c_2=0$, namely
 \beq 
H^{(0)}(z) = H_0 \left[\Om_{m0}(1+z)^3+\Om_\La \right]^{1/2},
\label{44d}
\eeq
and where we have defined $\Om_{m0}= 8 \pi G \rho_{m0}/3H_0^2$ and $\Om_\La= 8 \pi G \rho_{\La 0}/3H_0^2$. To first order in $c_2$ we then obtain  a simple algebraic equation for $H^{(1)}$, which can be easily solved and leads to:
\beq
H^{(1)}(z) = - {27\over 4} H_0^3 \Om_{m0}^2 (1+z)^6 \left[\Om_{m0}(1+z)^3+\Om_\La \right]^{-1/2}. 
\label{44e}
\eeq

In order to give to $\Om_{m0}$ the same physical meaning as in the standard scenario (namely, to interpret $\Om_{m0}$ as the present fraction of pressureless matter in units of the today value of the critical energy density, $\rho_c(t_0)= 3H^2(t_0)/ 8 \pi G$), we have now to impose that $H_0$ represents the today value of our generalized Hubble parameter, namely that
\beq
H_0= \lim_{z \to 0} H(z) =  \lim_{z \to 0} \left[ H^{(0)}(z)+ c_2 H^{(1)}(z)  \right].
\label{44f}
\eeq
By applying this condition to Eqs. (\ref{44d}) and (\ref{44e}) we obtain that the four parameters $H_0$, $c_2$, $\Om_{m0}$ and $\Om_\La$ cannot be independent, but must be related (to  first order in $c_2$) by the non-standard condition
\beq
\Om_\La=1 - \Om_{m0}+ {27\over 2} c_2 H_0^2 \Om_{m0}^2 ~.
\label{44g}
\eeq
By inserting this condition into Eqs. (\ref{44d}), (\ref{44e}), and combining all zeroth-order and first-order contributions to $H$, we can eventually write our approximate solution as follows:
\bea
&&
\!\!\!\!\!\!\!
H(z)= H^{(0)}+ c_2 H^{(1)}= H_0 \left[\Om_{m0}(1+z)^3+1- \Om_{m0} \right]^{1/2}
\nonumber \\ &&
\times
\left\{1+ {27\over 4} c_2 H_0^2 \Om_{m0}^2{1-(1+z)^6\over [\Om_{m0}(1+z)^3+1- \Om_{m0}}\right\}. 
\label{44h}
\eea
Let us stress again that the parameters $H_0$ and $ \Om_{m0}$ represent the same physical quantities as in the standard $\La$CDM scenario (namely, the today value of the Hubble parameter and of the critical fraction of pressureless matter). However, they are related to the critical fraction of dark energy $\Om_\La$ by the non-standard relation (\ref{44g}), imposed (to first order in $c_2$) by  our generalized cosmological equation (\ref{35}). 

Given the above (approximated) expression for $H(z)$ it is finally useful, for the application of this paper, to explicitly evaluate also the modified friction parameter and luminosity distance of Eqs. (\ref{41}) and (\ref{42}). By using Eqs. (\ref{43}) and (\ref{44}) we easily obtain, to first order in $c_2$:
\beq
\da(z) \simeq c_2 (1+z) R^{\prime (0)} = - 9 c_2 (1+z)^3 H_0^2 \Om_{m0},
\label{450}
\eeq
and 
\bea
\!\!\!\!\!\!\!
\frac{d_{L}^{GW}}{d_{L}}(z) &\simeq& 1+ c_2 \left[R^{(0)}(0)- R^{(0)}(z)\right]=
%\nonumber \\ 
%&=& 
1+ 3 c_2 H_0^2 \Om_{m0}\left[ (1+z)^3 -1 \right].
\label{451}
\eea

%%%%%%%%%%%%%%
%%%%%%%%%%%%%

\section {Numerical results using Supernovae data}
\label{sec5}
%\setcounter{equation}{0}

%%%%%%%%%%%%%%
%%%%%%%%%%%%%

By inserting the explicit form of our modified Hubble parameter, Eq. (\ref{44h}), into the luminosity distance (\ref{23}), we can now fit the Supernovae data to obtain a numerical estimate of the standard cosmological parameters,  as well as of the new parameter $c_2$. 

The method is the same as the one already applied in \cite{cosmai}, with the important difference that in this paper we have included the standard dark-energy contribution into the lowest-order background solution. In addition, we recall that the fit presented in \cite{cosmai} was performed at fixed given value of $H_0$, using for $H_0$ the value suggested by Planck data. Here we  prefer to keep  $H_0$ as a free parameter (after all, the local value of $H_0$ in the redshift range typical of Supernovae, $z\laq 1$, might be different from the corresponding value at much larger scales of distance). However, we will perform the fit at constant given value of $\Om_{m0}$, using to this purpose the recent results of Planck+BAO data (see e.g. \cite{26}).

We have fitted, in particular, the UNION 2 data set \cite{amulla} consisting of 557 measurements of redshift/distance-modulus of Type Ia Supernovae, with redshift between $0$ and $1.5$. We have used for the theoretical curve the standard expression of the so-called distance modulus $\mu(z)$, with the  normalization usually adopted for SNIa and defined by
\beq
\mu(z,c_2,H_0)= 5 \log_{10}\left[ 10^5 d_L(z)\over 1 {\rm Mpc}\right],
\label{45}
\eeq
where $d_L$ is given by Eq. (\ref{23}) and $H(z)$ is given by the generalized solution (\ref{44h}). Given the experimental points $\mu^{\rm obs}(z_i) \pm \Da \mu(z_i)$, where $\Da \mu$ is the relative error of the distance modulus for the i-th Supernova, we have performed a standard $\chi^2$-analysis, with
\beq
\chi^2= \sum_{i=1}^{557} \left[\mu^{\rm obs}(z_i)- \mu(z_i, c_2, H_0)\over  \Da \mu(z_i)\right]^2.
\label{46}
\eeq
By inserting into Eq. (\ref{44h}) the standard result \cite{26} $\Om_{m0} = 0.311 \pm 0.006$, and minimizing the above $\chi^2$ expression, we have then obtained the following best-fit values:
\bea
\!\!\!\!\!\!\!\!
H_0&=&(69.84 \pm 0.30) \,{\rm km \,\,s^{-1} Mpc^{-1}},
\nonumber \\ 
c_2&=&(1.128\pm 0.670)\times 10^{-6}  \,{\rm km^{-2} \,s^{2} \,Mpc^{2}},
\label{47}
\eea
with a goodness of fit $\chi^2/{\rm d.o.f} = 0.98$. 
Correspondingly, the critical fraction of dark energy turns out to be fixed, according to Eq. (\ref{44g}), to the value $\Om_\Lambda =0.696 \pm 0.100$. 
Note that the above value of $c_2$ is compatible with possible constraints arising from other cosmic data \cite{Ce}.  

 %%%%%%%%%%%%%%%
\begin{figure}[t]
\includegraphics[width=\columnwidth]{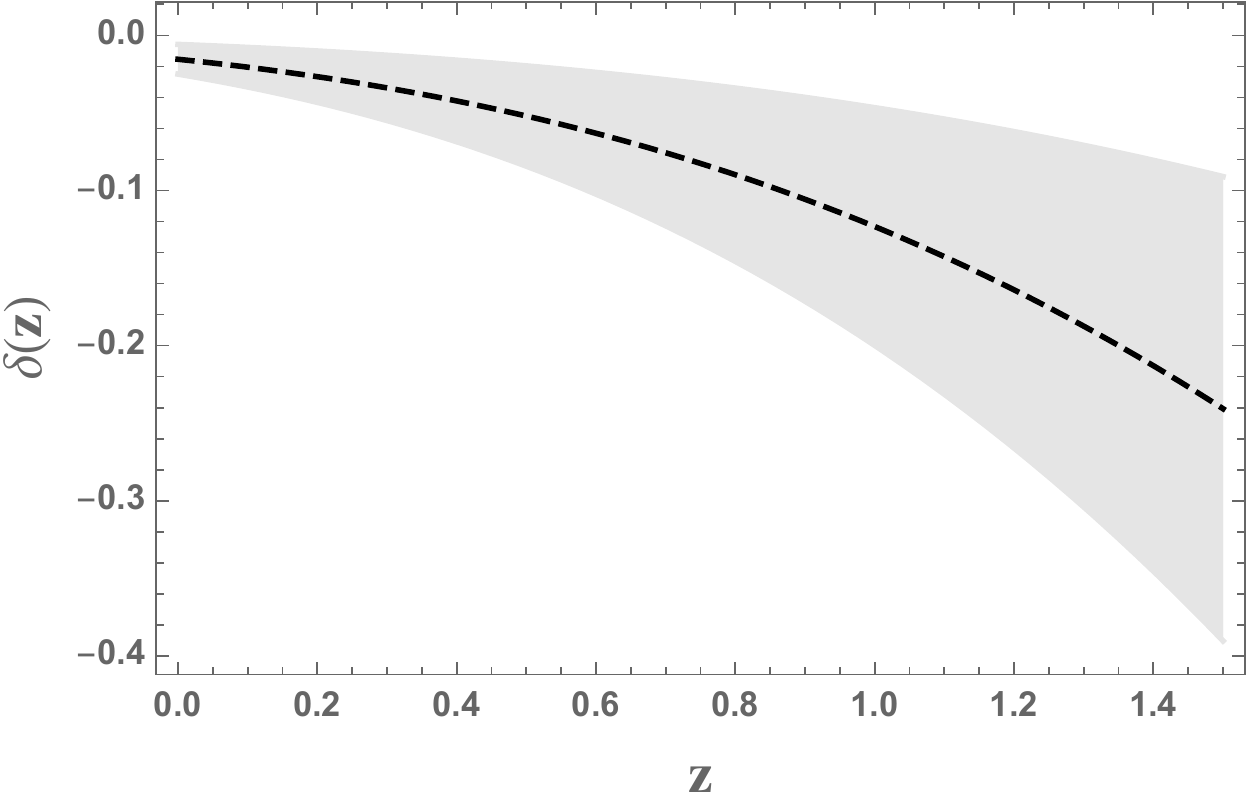}
\caption{Plot of the friction parameter $\da(z)$ of Eq. (\ref{450}) (black dashed curve), with the associated error band (grey region) controlled by the best-fit parameters of Eq. (\ref{47}). } 
\label{f1}
\end{figure}
%%%%%%%%%%%%%%%%

\begin{figure}[t]
\includegraphics[width=\columnwidth]{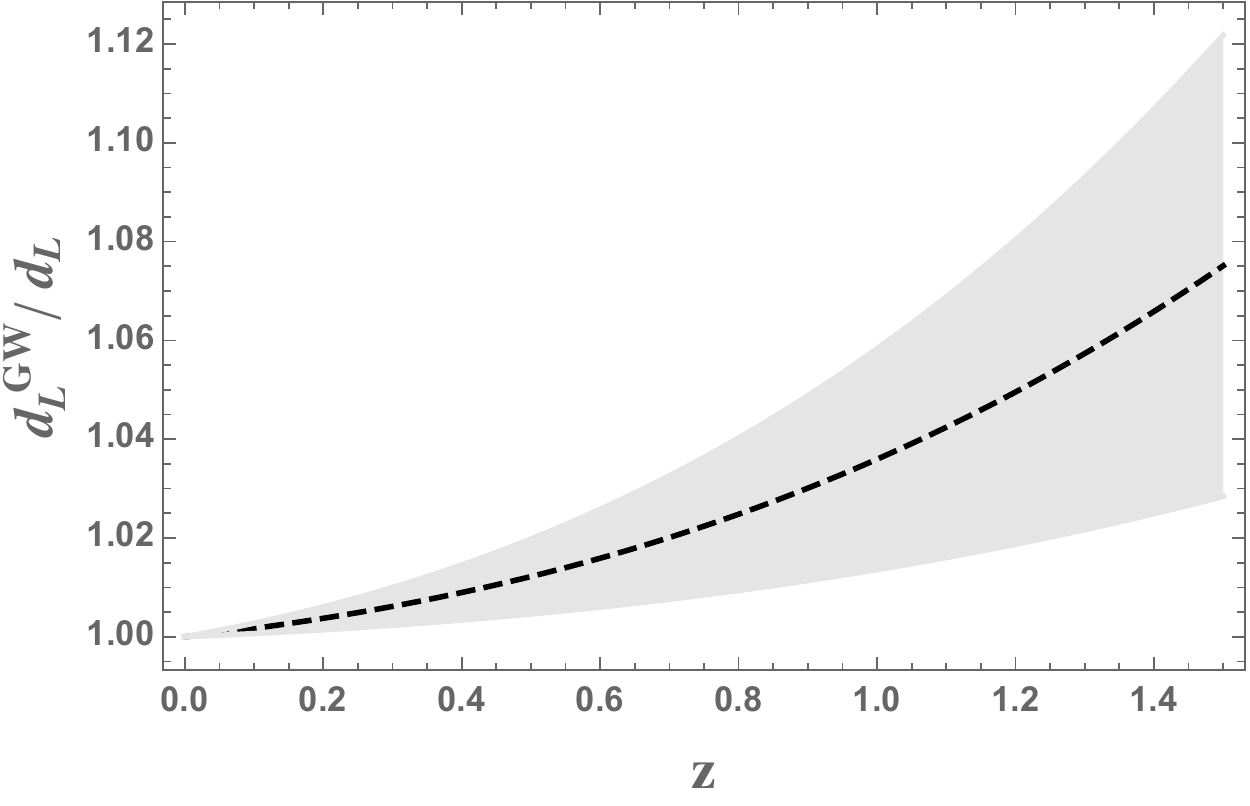}
\caption{Plot of the ratio (\ref{451}) between GW and e.m. luminosity distance (black dashed curve), with the associated error band (grey region) controlled by the best-fit parameters of Eq. (\ref{47}). } 
\label{f2}
\end{figure}

With the numerical values of Eq. (\ref{47}) we are now in the position of plotting the friction term (\ref{450}) and the modified GW luminosity distance (\ref{451}), consistently with present observational results on the chosen scale of distances. The results are shown in Figs. \ref{f1} and \ref{f2}, with the associated error bands obtained by propagating into Eqs. (\ref{450}) and (\ref{451}) the one sigma best-fit errors of Eq. (\ref{47}). 
It should be mentioned that we shall henceforth evaluate all errors under the conservative assumption that the uncertainties on the given parameters are to be summed with their absolute value. For the particular case in which there are no cross-correlations among the errors of the single parameters we are certainly overestimating the resulting error bars. However, the difference turns out to be very small and, in any case, our choice of using upper bound errors does not forbid a significant comparison with observations (and with standard predictions).

As shown by the plot of $\da(z)$ in Fig. \ref{f1}, the deviations from the standard Einstein predictions tend to increase with the distance, but they are very small in the limit  $z \ra 0$, where we find
\beq
\da(0)\simeq  - 0.015 \pm 0.009. 
\label{48}
\eeq
From the plot of Fig. \ref{f2} we can check the trivial identity $d_L^{GW}(0)= d_L(0)$ at $z=0$, consistently with the chosen normalization of the modified scale factor (see Eqs. (\ref{27}) and (\ref{28})). In the limit of large $z$ we can see that the modification of the luminosity distance tends to grow monotonically, as expected (at least in the considered range of distances), and it is important to note that $d_L^{GW} > d_L$ at all $z$. This means that, in the model of gravity we are considering, the received GW signals turn out to be {\em weaker} with respect to the value predicted by the standard Einstein theory (and thus
more difficult to be detected, given the same distance and detection sensitivity). 

 %%%%%%%%%%%%%%%
%%%%%%%%%%%%%%%%

Finally, comparing our results with those based on the gravitational model discussed in \cite{Belgacem:2017ihm,Belgacem:2018lbp} (the so-called RR model \cite{50,51}), it may be important to note that we find a different qualitative behavior for $\da$ and $d_L^{GW}$ as a function of $z$ (actually, the opposite behavior).  

In the RR model (in particular, in its minimal version where one of its parameters is set to zero) one finds  $\da(0) \simeq 0.062$, which is definitely larger than our result (\ref{48}). In the large $z$ limit, however, the ratio $d_L^{GW} /d_L$ tends to stabilize on the value $\simeq 0.970$ (see e.g. Fig. 2 and 3 of \cite{Belgacem:2018lbp}), which corresponds to a deviation from the standard Einstein result of about  $3 \%$. In our case, instaed, we expect a deviation of about $8 \%$ already at $z=1.5$ (see Fig. \ref{f2}).  It should be taken into account, however, that our plots and the plots of \cite{Belgacem:2018lbp} are performed not only with a different functional dependence on $z$, but also with a different $H_0$ parameter.

%%%%%%%%%%%%%%%%%%%%%%%%%%%%%%%%%%%%%%%%%%%%%%%%%%%%%%%%%%%%%%%%%%%%%%%%%%%%%%%%%%%
\subsection{Comparison with GW170817 data}
\label{sec5a}

It should be checked, at this point, whether the results we have presented for $\da$ and $d_L^{GW}$ may be compatible or not with other numerical estimates of such parameters, obtained from presently available data. Let us consider, in particular, the constraints on modified GW propagation following from the observation of the gravitational source GW170817 \cite{abbott6} and of its e.m. partner GRB170817A \cite{Goldstein:2017mmi}-\cite{GBM:2017lvd}.

In that case the redshift parameter is fixed at $z=0.00980$ (with an error which is negligible with respect to other uncertainties), and since this value is very small, we can use Eq. (\ref{28}) expanded to first order as
\beq
\frac{d_{L}^{GW}}{d_{L}}(z)\simeq 1-z\delta (0) + \cdots ~.
\label{49}
\eeq
Following the discussion of \cite{Belgacem:2018lbp}, we can then compare our previous result for $\da(0)$ with the ones obtained from the standard siren GW170817  in two ways: by extracting $\da(0)$ from the measured value of the Hubble parameter or from the measured luminosity distance.

Concerning the first approach, what is needed is the local Hubble constant $H_0^{GW}$, obtained from GW170817/GRB170817A measurements (updated to take into account the correct inclination of the source and the peculiar velocity of the host galaxy \cite{abbottnature,Guidorzi,abbott2019}), together with the local value of $H_0$ obtained with the e.m. measurements of a set of standard candles at very small redshift \cite{riess,riess2}. By using such values (reported in \cite{Belgacem:2018lbp}), by expanding to first order the luminosity-redshift relation as $d_L \simeq z/H_0+ \cdots$, by identifying, to this order, the measured e.m. distance with our $d_L$, one then obtains from Eq. (\ref{49}) the same result already presented in \cite{Belgacem:2018lbp}:
\beq
\delta (0)\simeq\frac{H_0^{GW}-H_0}{zH_0^{GW}}\simeq 2.7^{+15.4}_{-12.8}. 
\label{410}
\eeq
As expected, this is not a very stringent constraint, but it is well compatible with our previous estimate (\ref{48}).

Conversely, we may note that using our result (\ref{48}) for $\da(0)$, and the previously quoted \cite{riess,riess2} local value $H_0= (73.48
\pm 1.66)$ km s$^{-1}$ Mpc$^{-1}$ for the e.m. determined Hubble constant, we obtain from Eq. (\ref{410}):
\beq
H_0^{GW} \simeq H_0 \left[1+z \da(0)\right] \simeq (73.47
\pm 1.67) \,{\rm km \,\,s^{-1} Mpc^{-1}}.
\label{411}
\eeq
This agrees, within  the experimental errors, with the estimate of $H_0^{GW}$ obtained from the updated GW170817/GRB170817A measurements \cite{abbottnature,Guidorzi,abbott2019}, and given by $H_0^{GW}= 75.50^{+11.60}_{-9.60}$ km s$^{-1}$ Mpc$^{-1}$.

Finally, let us consider the second approach based on the direct comparison of the gravitational luminosity distance $d_L^{GW}$ of the standard siren GW170817 \cite{abbott6} and the e.m. luminosity distance $d_L^{em}$ of the host galaxy NGC4993 \cite{friedman,cantiello}. By working as before to lowest order in $z$, by identifying to this order $d_L^{em}$ with our $d_L$, and using the expression (\ref{49}), we are led to the numerical result also reported in \cite{Belgacem:2018lbp},
\beq
\delta (0)\simeq \frac{d_L-d_L^{GW}}{zd_L} \simeq
-7.8^{+9.7}_{-18.4}~~.
\label{412}
\eeq
Again, this is not a stringent limit, and is clearly compatible with our previous estimate (\ref{48}). 

Conversely, using the numerical value (\ref{48}) for $\da(0)$, using for $d_L$ the e.m. luminosity distance of the host galaxy \cite{friedman,cantiello},  $d_L= (40.7\pm2.4)$ Mpc, and inverting Eq. (\ref{412}),  we find no predicted difference for $d_L^{GW}$, namely $d_L^{GW}\simeq d_L$ . This is because of the very small deviations from the standard theory induced by our modified model of gravity in the small redshift limit. This results also agrees with the estimated gravitational distance of the standard siren \cite{abbott6}, $d_L^{GW} = 43.8^{+2.9}_{-6.9}$ Mpc.

%%%%%%%%%%%%%%%%%%%%%%%%%%%%%%%%%%%%%%%%%%%%%%%%%%%%%%%%%%%%%%%%%%%%%%%%%%%%%%%%%%%
\subsection{Comparison with the Einstein cosmological equations}
\label{sec5b}
%%%%%%%%%%%%%%%%%%%%%%%%%%%%%%%%%%%%%%
%%%%%%%%%%%%%%%%%%%%%%%%%%%%%%%%%%%%%%

In the standard cosmological scenario, based on the equations of the General Relativity theory, the luminosity distance of a source at redshift $z$, embedded in a spatially flat FLRW geometry, is given by the well-known expression:
\beq
d_{L}^{GR}(z) = \frac {1+z} {H_0} \int^z_0 \frac {dz'} {\left[\Omega_{m0} (1+z')^3+ (1-\Omega_{m 0})\right]^{1/2}}
\label{414} 
\eeq
(where GR denotes General Relativity). 
It depends on two observational parameters, $H_0$ and $\Om_{m0}$, and is the same for both e.m. and GW signals (in general, it is the same for all signals propagating on the light-cone of the given FLRW metric background).

In a cosmological scenario based on the modified gravitational equations (\ref{32}) we have instead two different types of luminosity distance. 

We have still the luminosity distance $d_L^{em}$ associated with light-cone propagation (and typical, for instance of e.m. signals): it is defined by Eq. (\ref{23}), where, however, the function $H(z)$ is now a solution of the generalized gravitational equations and is thus different, in general, from the standard expression of $H(z)$ appearing in Eq. (\ref{414}). In addition, we have a modified luminosity distance $d_L^{GW}$ associated with the propagation of GW signals (distorted by the presence of a non standard friction coefficient), and related to $d_L^{em}$ by Eq. (\ref{42}). The aim of this Subsection is to compare the generalized expressions of both $d_L^{em}$ and $d_L^{GW}$  with the standard luminosity distance $d_L^{GR}$  of Eq. (\ref{414}).

To this purpose, and for  a more consistent comparison, let us first evaluate the parameters controlling the standard expression of  $d_L^{GR}$  in the same way used for our model of modified gravity: namely, in the same redshift range, by fitting the same experimental data as before, the UNION 2 data set \cite{amulla}, 
and at the same fixed value of the parameter $\Om_{m0}$, namely \cite{26}  $\Om_{m0}=0.311 \pm 0.006$. The standard best-fit procedure, starting with the expression (\ref{45}) of the distance modulus written for the luminosity distance (\ref{414}), then gives:
\beq
H_0=(69.53 \pm 0.24) \,{\rm km \,\,s^{-1} Mpc^{-1}},
\label{415}
\eeq
with $\chi^2/{\rm d.o.f}=0.98$. This completely fixes (with the relative uncertainty) the standard model luminosity (\ref{414}) as a function of $z$. 

\begin{figure}[t]
\includegraphics[width=\columnwidth]{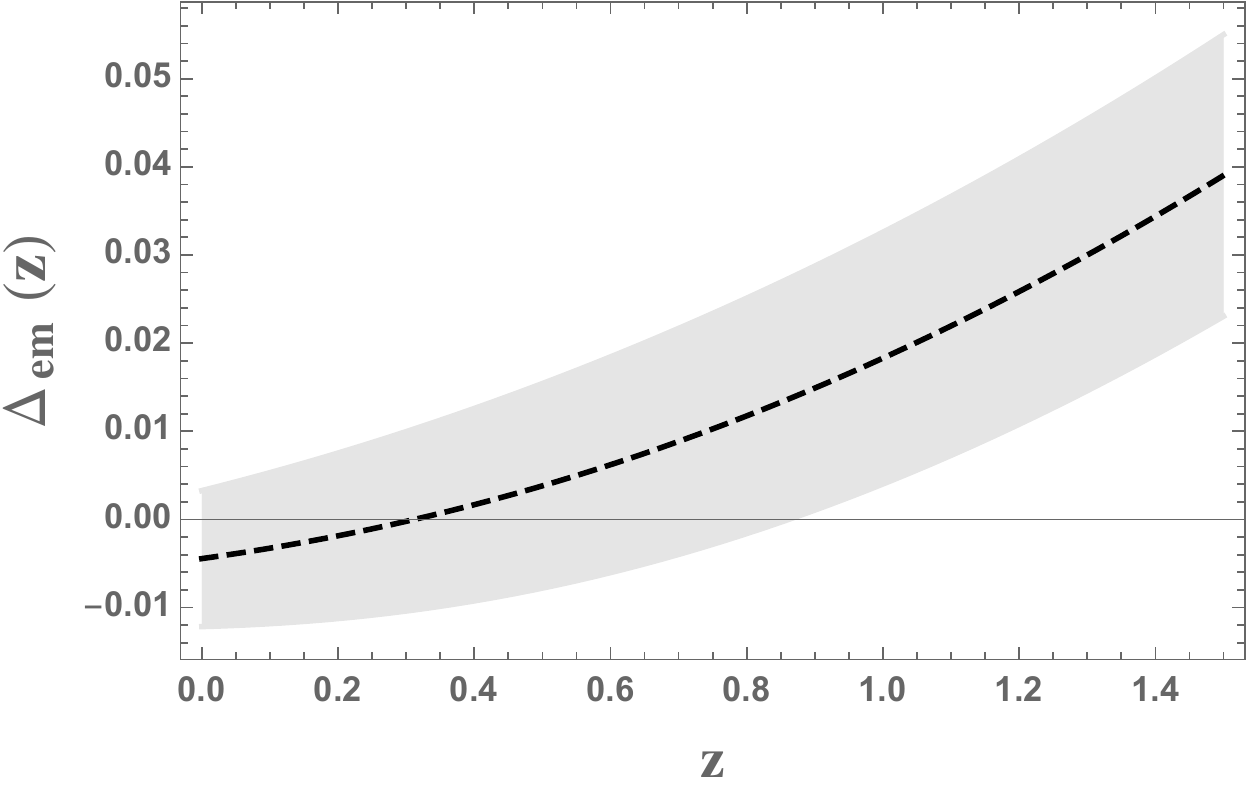}
\caption{Plot of $\Da_{em}(z)$ (black dashed curve) with the relative error band (grey region), controlled by the best fit parameters of Eqs. (\ref{415}) for $d_L^{GR}$ and of Eq. (\ref{47}) for $d_L^{em}$. In both cases we have set $\Om_{m0}=0.311 \pm 0.006$. The use of different parameters for the two luminosity distances  leads to $\Da_{em}(0)\not=0$ (see the text).} 
\label{f3}
\end{figure}

\begin{figure}[ht]
\includegraphics[width=\columnwidth]{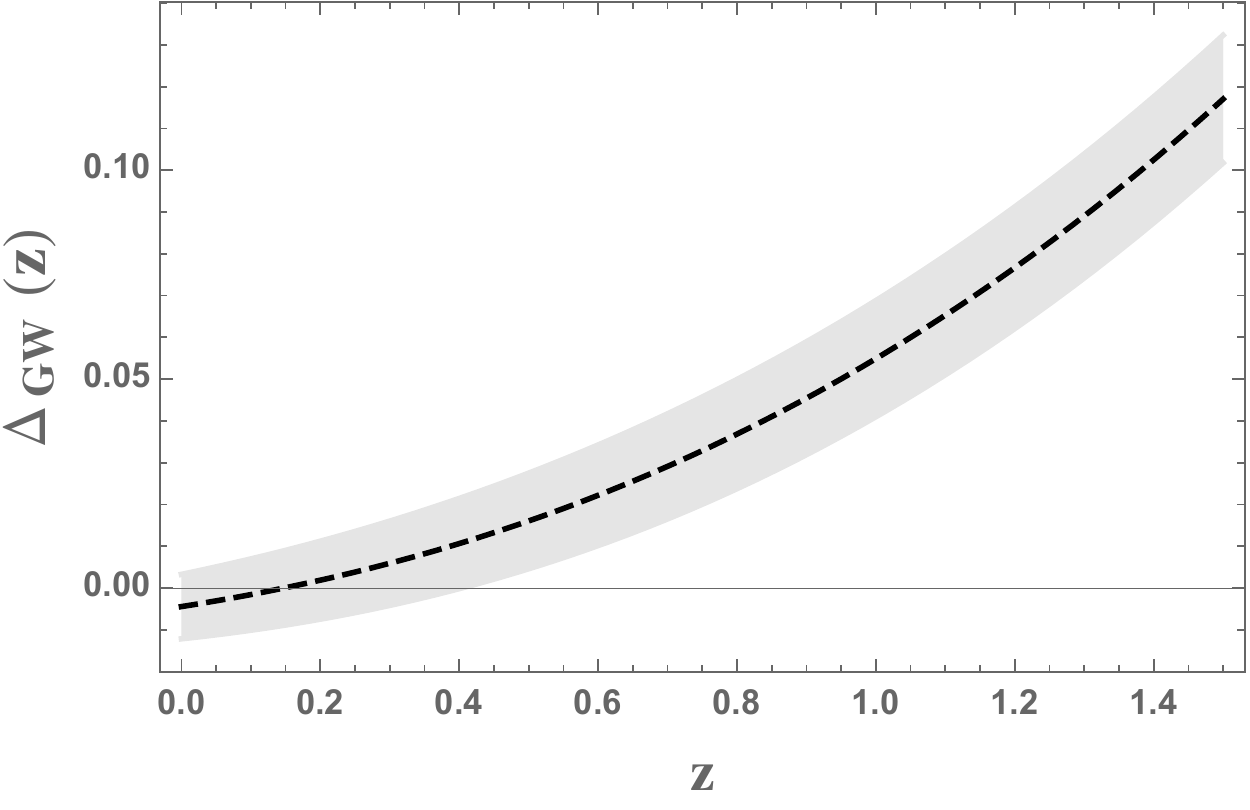}
\caption{Plot of $\Da_{GW}(z)$ (black dashed curve) with the relative error band (grey region), controlled by the best fit parameters of Eqs. (\ref{415}) for $d_L^{GR}$ and of Eq. (\ref{47}) for $d_L^{GW}$. In both cases we have set $\Om_{m0}=0.311 \pm 0.006$. The use of different parameters for the two luminosity distances leads to $\Da_{GW}(0)\not=0$ (see the text).} 
\label{f4}
\end{figure}

We are now in the position of comparing the above expression of $d_L^{GR}$ with the e.m. distance $d_L^{em}$ of our model -- given by Eq. (\ref{23}) in terms of the generalized $H(z)$ of Eq. (\ref{44h}) -- and with the GW distance $d_L^{GW}$,  related to $d_L^{em}$ by Eq. (\ref{451}). 
Of course, the modified luminosity distances are to be computed (as emphasized also in \cite{Belgacem:2018lbp}) with the parameters determined by the associated fitting procedures (i.e. the ones reported in  Eq. (\ref{47})), while the standard distance (\ref{414}) refers to the parameter (\ref{415}).
The plots of the corresponding fractional corrections,
\bea
&&
\Da_{em}(z)= { d_L^{em}- d_L^{GR}\over d_L^{GR}},
~~~~~~~~~
%\nonumber \\ &&
\Da_{GW}(z)= { d_L^{GW}- d_L^{GR}\over d_L^{GR}},
\label{416}
\eea
together with the associated error bars, are presented in Fig. \ref{f3} and \ref{f4}. 

 %%%%%%%%%%%%%%%
%%%%%%%%%%%%%%%%
% %%%%%%%%%%%%%%%
%%%%%%%%%%%%%%%%

%%%%%%%%%%%%%%%%%%%%%%%%%%%%%%%%%%%%%%
%%%%%%%%%%%%%%%%%%%%%%%%%%%%%%%%%%%%%%

We can easily check that the behavior of the plotted  functions is in agreement (at all $z$) with the property $d_L^{GW}>d_L^{em}$, already emphasized by Fig. \ref{f2} and due, as previously stressed, to a faster decrease with the distance of the GW signals with respect to e.m. signals. We can also see, from Fig. \ref{f3}, that $\Da_{em}$ tends to be positive at large enough $z$, and this implies that also the e.m. signals, in our modified-gravity scenario, tend to be weaker than in the standard scenario, at large enough distances. In addition, by comparing the plots of Fig. \ref{f3} and Fig. \ref{f4}, we find that $|\Da_{GW}|>|\Da_{em}|$, at any $z$ (at least in the considered redshift range). This seems to suggest that detecting GW signals might represent a more efficient tool (than detecting e.m. signals) to test generalized model of gravity on the given scale of cosmic distances.

Finally, it is important to recall that in the limit $z \ra 0$ the two modified distances tend (by definition) to coincide, i.e. $d_L^{em}(0)=d_L^{GW}(0)$ (as previously stressed, and as illustrated by Fig. \ref{f2}). This explains why, for $z=0$, the two fractional corrections of 
Fig. \ref{f3} and Fig. \ref{f4} go to the same (non-vanishing) value
\beq
\Da_{em}(0)=\Da_{GW}(0)\simeq -0.004 \pm 0.008.
\label{417}
\eeq

Note that the difference $\Da(0)$ (still compatible with zero, to one sigma, in the context of our numerical analysis), in principle can always be expected to be non-vanishing, because we have plotted the standard and the modified luminosity distances using, for each curve, their peculiar best-fit parameters: that of Eq. (\ref{415}) for the Einstein gravitational equations, and those of Eq. (\ref{47}) for the modified equations. Let us note, in this respect, that the importance of using for each model its own set of estimated parameters has been repeatedly stressed also in 
\cite{Belgacem:2018lbp}. Only in this way we  obtain for $\Da_{em}(z)$ and $\Da_{GW}(z)$ a relevant quantity to be compared with observations \cite{Belgacem:2018lbp}. This implies, however, that the results we are presenting cannot be reliably extrapolated outside the range of $z$ that we have considered to fit the data, and to obtain the corresponding estimates.

%%%%%%%%%%%%%%%%%%%%%%%%%%%%%%%%%%%%%%
%%%%%%%%%%%%%%%%%%%%%%%%%%%%%%%%%%%%%%

\section {Conclusion}
\label{sec6}

In this paper we have studied the possible differences arising between the propagation of GW signals and e.m. signals -- as well as the differences between the related luminosity distances -- in the context of a generalized theory of gravity with quadratic curvature corrections. 

Considering in particular the range of distances typical of Supernovae ($z \laq 1.5$), we have found that the deviations from the standard expression of the luminosity distance is larger for GW signals than for e.m. signals (at least, in the range of redshift we have considered). This seems to confirm previous suggestions \cite{Belgacem:2018lbp,B1} that GW detectors might play a crucial (future) role in discriminating between different models of gravity on cosmic scales of distances. For the case discussed in this paper, however, the amplitude of   GW signals received by a distant source, after their modified propagation, tends to be weaker than in the context of the standard Einstein theory. 

It should be stressed, finally, that the main purpose of this paper is not only to propose possible observational constraints on the particular model of gravity we have considered, but also to present and discuss a general approach, possibly useful for future applications to other (more general and motivated) non-standard theories of gravity, as well as to other types of astrophysical signals (like neutrinos, whose propagation is also crucially affected by the given background geometry \cite{DSG}). 

The particular choice of the model discussed in this paper has been motivated by previous applications of the same model in an astrophysical context to study deviations from the standard cosmological dynamics \cite{43}, as well as the detailed mechanisms of neutron star formations \cite{44,45}. The validity of the analysis presented in this paper is limited to the redshift range $z \laq 1.5$. We are planning to discuss the compatibility of this model with other available cosmological data, and in a larger redshift range, in a future paper.

%%%%%%%%%%%%%%%%%%%%%%%%%%%%%%%%%%%%%%
%%%%%%%%%%%%%%%%%%%%%%%%%%%%%%%%%%%%%%

\paragraph{\bf{ Appendix}}
\setcounter{equation}{0}
\renewcommand\theequation{A.\arabic{equation}}

Let us consider the field equations (\ref{32}) for our generalized model of gravity, and expand to first order the metric tensor, $g_{\mu\nu} \ra g_{\mu\nu}+h_{\mu\nu}$, around the FLRW  background solution described by $g_{00}= 1$, $g_{ij}= -a^2(t) \da_{ij}$. Let us work in the TT gauge where $h_{\mu0}=0$, $\nabla_jh_i\,^j\equiv \pa_j h_i\,^j=0$, $g^{ij}h_{ij}=0$. In such a context we have the useful identities: 
\bea
&&
g^{kj}\dot{h}_{ik}=\dot{ h_{i}}^{j}+2H{ h_{i}}^{j},
\label{a1} \\ &&
g^{kj}\ddot{h}_{ik}=
\ddot{ h_{i}}^{j}+2\dot{H}{ h_{i}}^{j}+4H\dot{ h_{i}}^{j}+4H^{2}{ h_{i}}^{j},
\label{a2}
\eea
and we find $\da R_{00}=0$, $\da R_{0i}=0$,
\bea
&&
\!\!\!\!\!\!\!\!
\delta R_{ij}=-\frac{1}{2}\ddot{h}_{ij}+\frac{\nabla^{2}}{2a^{2}}h_{ij}+\frac{1}{2}H\dot{h}_{ij}-2H^{2}h_{ij},
\label{a3} \\ && \!\!\!\!\!\!\!\!
\da R= -h^{ij} R_{ij}+g^{ij} \da R_{ij} \equiv 0.
\label{a4}
\eea
The first-order perturbation of the field equation (\ref{32}) thus reduces to
\beq
\delta{R_{\mu}}^{\nu}+2c_{2}\left[R\delta{R_{\mu}}^{\nu}+{\delta_{\mu}}^{\nu}\delta (\Box R)-\delta( \nabla_{\mu}\nabla^{\nu}R)\right]=0,
\label{a5}
\eeq
where
\bea
&&
 \delta{R_{0}}^{0}=0=\delta{R_{i}}^{0}, ~~~ ~~~~~
 %\nonumber \\ &&
{\delta R_{i}}^{j}=-\frac{1}{2}\left(\ddot{ h_{i}}^{j}+3H\dot{ h_{i}}^{j}- \frac{\nabla^{2}}{a^{2}}{ h_{i}}^{j}\right)
\label{a6}
\eea
(we have used Eq. (\ref{a2}) to rewrite the Ricci curvature perturbation (\ref{a3}) in the form of Eq. (\ref{a6})). 

We are left with the perturbation of the higher-derivative terms $\nabla_\mu\nabla_\nu R$. Using Eq. (\ref{a4}) we obtain
\bea
&&
\da (\nabla_0\nabla_0 R)= \da (\nabla_0\nabla_i R)=0, ~~~~~~~~~
%\nonumber \\ &&
\delta\left(\nabla_{i}\nabla_{j}R\right)=\frac{\dot{R}}{2}\,\dot{h}_{ij},
\label{a7}
\eea
so that, using Eq. (\ref{a1}),
\beq
\delta (\Box R) = -h^{ij}\nabla_{i}\nabla_{j}R+g^{ij}\delta (\nabla_{i}\nabla_{j}R)=0.
\label{a8}
\eeq
Finally, by using Eqs. (\ref{a7}) and (\ref{a1}),
\beq
\delta (\nabla_{i}\nabla^{j}R)=-h^{jk} \nabla_{i}\nabla_{k}R+\frac{\dot{R}}{2}\,g^{jk}\dot{h}_{ik}=
\frac{\dot{R}}{2}\dot{ h_{i}}^{j}.
\label{a9}
\eeq

By inserting all contributions into Eq. (\ref{a5}) we find that the temporal and mixed components of such perturbed equations are identically satisfied, while the spatial components give us the modified propagation equations:
\beq
\ddot{ h_{i}}^{j}+\biggl(3H+\frac{2c_{2}\dot{R}}{1+2c_{2}R}\biggr)\dot{ h_{i}}^{j}-\frac{\nabla^{2}}{a^{2}}{ h_{i}}^{j}=0.
\label{a10}
\eeq
In conformal time, where $\dot h= h'/a$, the equation can be rewritten as
\beq
 h''_{i}\,^{j}+2{a'\over a}\left[1+{c_2R'\over {\cal H}(1+2c_2R)}\right] h'_{i}\,^{j}-\nabla^{2}h_{i}\,^{j}=0,
\label{a11}
\eeq
and thus defines the friction coefficient $\da(\eta)$ reported in Eq. (\ref{38}).

%%%%%%%%%%%%%%%%%%%%%%%%%
%%%%%%%%%%%%%%%%%%%%%%%%%

%\vspace{1cm}

\begin{acknowledgements}
Maurizio Gasperini and Luigi Tedesco are supported in part by INFN under the program TAsP ({\it ``Theoretical Astroparticle Physics"}), and by the research grant number 2017W4HA7S {\it ``NAT-NET: Neutrino and Astroparticle Theory Network"}, under the program PRIN 2017 funded by the Italian Ministero dell'Universit\`a e della Ricerca (MUR). Giuseppe Fanizza acknowledges support by FCT under the program {\it ``Stimulus"} with the grant no. CEECIND/04399/2017/CP1387/CT0026. 
It is a pleasure to thank Gabriele Veneziano for many discussions and useful suggestions.
\end{acknowledgements}




\end{document}